\documentclass[twocolumn,showpacs,preprintnumbers,amsmath,amssymb]{revtex4}

\usepackage{graphicx}
\usepackage{dcolumn}
\usepackage{bm}
\def\bk{\!\!\!/}

\begin{document}

\preprint{In memory of Ian Kogan}

\title{Vector meson couplings to vector and tensor currents\\
in extended NJL quark model}

\author{Mihail Chizhov}
\affiliation{Institut de Physique Nucl\'eaire de Lyon\\
and Universit\'e Claude Bernard Lyon-1\\
69622 Villeurbanne, France\\
and\\
Centre for Space Research and Technologies,
Faculty of Physics,\\ University of Sofia, 1164 Sofia, Bulgaria
}%

\date{\today}

\begin{abstract}
A simple explanation of the dynamic properties of vector mesons is given 
in the framework of extended Nambu--Jona-Lasinio quark model. 
New mass relations among the hadron vector resonances are derived. 
The results of this approach are in good accordance with the QCD sum rules, 
the lattice calculations and the experimental data.
\end{abstract}


\maketitle

\section{Introduction}

All known elementary vector particles, the photon, $Z$,$W$ and 
the gluons, are gauge particles. 
An opinion exists that there are no other vector particles besides the gauge 
ones. They have only chirality
conserved vector interactions with the matter fields. The so called
{\it anomalous} term, 
which appears from radiative corrections~\cite{schwinger}, 
is small and is not present in the initial Lagrangian. 
It describes chirality flipped interactions with the tensor current. 

Although the gluons are gauge particles, they induce chirality violation
and, therefore, in the hadron physics the {\it anomalous} interactions 
of the vector mesons are not so small. 
They should be taken into account, for example
to extract $|V_{ub}|$ element of CKM matrix from the semileptonic
decay $B\to\rho\ell\nu$. Even more peculiar is the existence of the vector 
$b_1(1235)$ meson with only anomalous interactions with quarks. 
It is not possible at all to describe this meson as a gauge particle.

Such kind of particles originate from a tensor formalism.
It is known that free relativistic particles with spin 1 can be described 
by the four-vector $A_\mu$ or by the second rank antisymmetric tensor
field $T_{\mu\nu}$. These two different descriptions are applied to the
vector mesons without noting the main difference between them~\cite{Ecker}.
However, the different formalisms are related to the different couplings
of the vector mesons to the quarks and, therefore, to the different chiral
properties. The vector fields are transformed under {\it real}
(chirally neutral) representation (1/2,1/2) of the Lorentz group. On the 
other hand $T_{\mu\nu}\pm i\tilde{T}_{\mu\nu}$ combinations are
transformed under irreducible {\it chiral} representations (1,0)
and (0,1). They describe {\it chiral} vector fields.

The aim of this letter is to put on the same footing
(at least at a phenomenological level) the consideration
of the gauge particles and the particles with {\it anomalous} only
interactions.
The successful description of the dynamical properties of the hadron systems
hints that the same phenomenon takes place in the
high energy physics too. Then we may expect the discovery of 
{\it anomalously} interacting particles at future colliders.

The Nambu--Jona-Lasinio (NJL)~\cite{NJL} quark model
is a successful tool for investigating hadron physics 
and the spontaneous chiral symmetry breaking mechanism.
However, an extension of the model is needed~\cite{ch1} in order 
to introduce the new type {\it anomalous} interactions
and the particles associated with them .

To simplify the idea we will deal only with the one-flavor NJL model.
The generalization for the real case with $N$ flavors is straightforward 
and will be considered elsewhere. As long as the isospin triplet $I=1$
consists of {\it up}
and {\it down} quarks with approximately the same constituent masses, 
we can apply this one-flavor model to the vector mesons
$\rho$, $b_1$, $a_1$ and $\rho'$. Using this approach we will get,
for example, 
very simple prediction for the ratio $f^T_\rho/f_\rho\simeq 1/\sqrt{2}$,
which matches very well the latest lattice calculations~\cite{lattice}.

\section{The model}

Following the classical paper~\cite{NJL} we start with the chiral invariant
Lagrangian
\begin{eqnarray}
{\cal L}=&&\bar{\psi}q\bk\psi+\frac{G_0}{2}~
\bar{\psi}\!\left(1+\gamma^5\right)\!\psi~
\bar{\psi}\!\left(1-\gamma^5\right)\!\psi\nonumber\\
&-&\frac{G_V}{2}\left(\bar{\psi}\gamma_\mu\psi\right)^2
-\frac{G_A}{2}\left(\bar{\psi}\gamma_\mu\gamma^5\psi\right)^2\nonumber\\
&-&\frac{G_T}{2}~\bar{\psi}\sigma_{\mu\lambda}\!\left(1+\gamma^5\right)\!\psi
\frac{q_\mu q_\nu}{q^2}
\bar{\psi}\sigma_{\nu\lambda}\!\left(1-\gamma^5\right)\!\psi
\label{4fermion}
\end{eqnarray}
where the new tensor interaction term was introduced. 
It should contain an unique
momentum dependence, because the local product of two tensor
currents with different chiralities vanishes identically.
Generally speaking, there are four different positive coupling parameters 
$G_0$, $G_V$, $G_A$ and $G_T$ with dimensions $[mass]^{-2}$
for each chiral invariant term.

Let us rewrite the Lagrangian (\ref{4fermion})
by introducing auxiliary bosonic fields (without kinetic terms)
which will later play the role of collective meson states after
quantization
\begin{eqnarray}
{\cal L}=&&\!\bar{\psi}q\bk\psi
+g_\sigma\bar{\psi}\psi \sigma-\frac{g_\sigma^2}{2G_0}\sigma^2
+ig_\pi\bar{\psi}\gamma^5\!\psi \pi-\frac{g_\pi^2}{2G_0}\pi^2\nonumber\\
&+&\!g_V\bar{\psi}\gamma_\mu\psi V_\mu\!+\!\frac{g_V^2}{2G_V}V_\mu^2
+g_A\bar{\psi}\gamma_\mu\gamma^5\!\psi A_\mu\!+\!\frac{g_A^2}{2G_A}A_\mu^2
\nonumber\\
&-&\!ig_R~\bar{\psi}\sigma_{\mu\nu}\psi~\frac{q_\mu}{|q|}R_\nu
+\frac{g_R^2}{2G_T}R_\mu^2\nonumber\\
&+&\!g_B~\bar{\psi}\sigma_{\mu\nu}\gamma^5\!\psi~\frac{q_\mu}{|q|}B_\nu
+\frac{g_B^2}{2G_T}B_\mu^2,
\label{boson}
\end{eqnarray}
where $g_a$ ($a=\sigma,\pi,V,A,R,B$) are dimensionless coupling constants.
At the classical level the Lagrangians (\ref{4fermion}) and (\ref{boson})
are equivalent. However, the perturbation quantum field theory cannot
be applied in the former case due to the dimensional constants $G$. Therefore,
the second form of the Lagrangian with dimensionless coupling constants $g_a$
is more appropriate for quantization by perturbative methods~\cite{eguchi}.

\section{Quantum corrections and symmetry breaking}

\begin{figure}[h]
\begin{picture}(200,70)(10,20)
\multiput(16,55)(51,0){2}{$g_a$}
\put(45,50){\circle{40}}
\multiput(25,50)(40,0){2}{\circle*{2}}
\multiput(0,50)(10,0){3}{\line(1,0){5}}
\multiput(65,50)(10,0){3}{\line(1,0){5}}
\put(45,70){\vector(1,0){1}}
\put(45,30){\vector(-1,0){1}}
\put(42,15){(a)}
\put(155,50){\circle{40}}
\multiput(128,59)(48,0){2}{$g_\sigma$}
\multiput(128,38)(48,0){2}{$g_\sigma$}
\put(155,70){\vector(1,0){1}}
\put(175,50){\vector(0,-1){1}}
\put(155,30){\vector(-1,0){1}}
\put(135,50){\vector(0,1){1}}
\multiput(120,17)(4,4){6}{\circle*{2}}
\multiput(140,63)(-4,4){6}{\circle*{2}}
\multiput(170,63)(4,4){6}{\circle*{2}}
\multiput(170,37)(4,-4){6}{\circle*{2}}
\put(152,15){(b)}
\end{picture}
\caption{\label{rad}Radiative corrections to (a) meson self-energy and 
(b) scalar self-interaction.}
\end{figure}
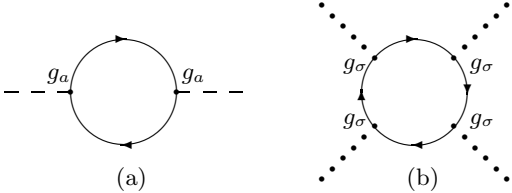

\begin{figure}[h]
\begin{picture}(200,70)(10,20)
\put(50,50){\circle{40}}
\multiput(23,59)(48,-21){2}{$g$}
\multiput(23,38)(48,21){2}{$g_\sigma$}
\put(50,70){\vector(1,0){1}}
\put(70,50){\vector(0,-1){1}}
\put(50,30){\vector(-1,0){1}}
\put(30,50){\vector(0,1){1}}
\multiput(15,17)(4,4){6}{\circle*{2}}
\multiput(32.5,63)(-2.5,2.5){7}{\oval(5,5)[t]}
\multiput(30,65.5)(-2.5,2.5){7}{\oval(5,5)[r]}
\multiput(65,63)(4,4){6}{\circle*{2}}
\multiput(67.5,32)(2.5,-2.5){7}{\oval(5,5)[t]}
\multiput(65,34.5)(2.5,-2.5){7}{\oval(5,5)[r]}
\multiput(35,63)(30,-26){2}{\circle*{2}}
\multiput(80,83)(-70,-66){2}{\line(1,0){10}}
\multiput(85,78)(-70,-66){2}{\line(0,1){10}}
\put(47,15){(a)}
\put(126,56){$g_V$}
\put(175,38){$g_R$}
\put(175,60){$g_\sigma$}
\put(155,50){\circle{40}}
\put(135,50){\circle*{2}}
\multiput(105,50)(8,0){4}{\oval(4,5)[t]}
\multiput(109,50)(8,0){4}{\oval(4,5)[b]}
\multiput(170,37)(0,26){2}{\circle*{2}}
\multiput(172.5,32)(2.5,-2.5){7}{\oval(5,5)[t]}
\multiput(170,34.5)(2.5,-2.5){7}{\oval(5,5)[r]}
\put(145.5,32.5){\vector(-2,1){1}}
\put(145.5,67.5){\vector(2,1){1}}
\put(175,50){\vector(0,-1){1}}
\multiput(170,63)(4,4){6}{\circle*{2}}
\put(185,83){\line(1,0){10}}
\put(190,78){\line(0,1){10}}
\put(152,15){(b)}
\end{picture}
\caption{\label{breaking} Contributions to the mass term (a) 
and mixing between vector mesons (b).}
\end{figure}
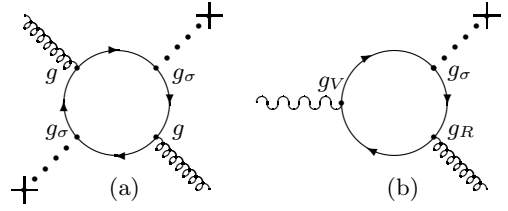

It is interesting to note that in spite of absence of kinetic terms 
for the meson fields in the initial Lagrangian~(\ref{boson}) 
they are generated
on the quantization stage due to radiative corrections (Fig.~\ref{rad}a).
Such kinetic terms generation reflects into relations of the
coupling constants $g_a$. In one-loop approximation they are
\begin{equation}
3g_\sigma^2=3g_\pi^2=2g_V^2=2g_A^2=g_R^2=g_B^2=\frac{24\pi^2}{N_c}~\varepsilon,
\label{ga}
\end{equation}
where $N_c$ is the number of colors and $\varepsilon$ is the 
dimensional regularization parameter.
The quark loops lead also to various self-interactions and   
interactions among the mesons.
For example, due to generation of quartic self-interaction $\sigma^4$ 
of the scalar meson (Fig.~\ref{rad}b), 
a spontaneous breaking of the chiral symmetry occurs
which leads to generation of non-zero quark mass
\mbox{$m=-\langle\sigma\rangle/g_\sigma$}, to additional mass 
contributions for the mesons (Fig.~\ref{breaking}a) 
and to mixings between them (Fig.~\ref{breaking}b).

The chiral symmetry breaking for the scalar sector has been studied in detail.
The introduction of the new vector mesons $R_\mu$ and $B_\mu$ brings
nothing new there. On the other hand the symmetry breaking leads to an
interesting phenomenon in the vector mesons sector. In the
following we focus just on the vector mesons.

We have introduced four vector mesons $V_\mu$, $A_\mu$, $R_\mu$ and $B_\mu$.
Their quantum numbers (the total angular momentum $J$, $P$-parity and 
the charge conjugation $C$)
can be defined from their interactions with quarks (\ref{boson}):
\begin{center}
\begin{tabular}{l|c|c|c|c}
meson vector fields:    & $V_\mu$ & $A_\mu$ & $R_\mu$ & $B_\mu$\\
\hline
quantum numbers $J^{PC}$: & $1^{--}$ & $1^{++}$ & $1^{--}$ & $1^{+-}$
\end{tabular}
\end{center}
The vector mesons $V_\mu$ and $A_\mu$ have gauge-like 
minimal interactions
with quarks, while $R_\mu$ and $B_\mu$ have only
{\it anomalous} tensor interactions. Nevertheless,
in the chiral limit all these mesons couple to {\it conserved} quark currents.
Therefore, their kinetic terms are gauge invariant 
\begin{eqnarray}
{\cal L}_0=\!\!\!&&-\frac{1}{2}V_\mu
\!\left(q^2g_{\mu\nu}-q_\mu q_\nu\right)V_\nu +\frac{M^2_V}{2}V^2_\mu
\nonumber\\
\!\!\!&&-\frac{\sqrt{18}m}{|q|}~V_\mu
\!\left(q^2g_{\mu\nu}-q_\mu q_\nu\right)R_\nu\nonumber\\
\!\!\!&&-\frac{1}{2}R_\mu\!\left(q^2g_{\mu\nu}-q_\mu q_\nu\right)R_\nu
+\frac{M^2_T-6m^2}{2}R^2_\mu\nonumber\\
\!\!\!&&-\frac{1}{2}B_\mu\!\left(q^2g_{\mu\nu}-q_\mu q_\nu\right)B_\nu
+\frac{M^2_T+6m^2}{2}B^2_\mu\nonumber\\
\!\!\!&&-\frac{1}{2}A_\mu\!\left(q^2g_{\mu\nu}-q_\mu q_\nu\right)A_\nu
+\frac{M^2_A+6m^2}{2}A^2_\mu.
\label{kin}
\end{eqnarray}
Here $M^2_V=g^2_V/G_V$, $M^2_A=g^2_A/G_A$ and 
$M^2_T=g^2_R/G_T=g^2_B/G_T$
are the initial vector boson masses before symmetry breaking. Since  
$R_\mu$ and $B_\mu$ are chiral mesons, their masses $M^2_R=M^2_T-6m^2$
and $M^2_B=M^2_T+6m^2$ split due to
symmetry breaking. The axial-vector mesons $A_\mu$ also get additional 
contribution to their mass terms.

The essential feature of eq.\ (\ref{kin}) is the mixing between the two
vector mesons $V_\mu$ and $R_\mu$ which have the same quantum numbers
as $\rho$ and $\rho'$ mesons. In the Lorentz gauge this reads 
\begin{equation}
{\cal L}_{VR}=-\frac{1}{2}\left(V_\mu R_\mu\right)
\left(\begin{array}{cc}
q^2-M^2_V & \sqrt{18}m|q|\\
\sqrt{18}m|q| & q^2-M^2_R
\end{array}\right)\left(\begin{array}{c}
V_\mu\\R_\mu
\end{array}\right)
\label{VR}
\end{equation}
If we identify the physical states with quantum numbers $1^{--}$ 
with these mesons, they can be expressed as linear
combinations of the chiral eigenstates $V_\mu$ and $R_\mu$ 
\begin{eqnarray}
\rho_\mu(q^2)&=&~~\cos\theta(q^2)V_\mu+\sin\theta(q^2)R_\mu\nonumber\\
\rho'_\mu(q^2)&=&-\sin\theta(q^2)V_\mu+\cos\theta(q^2)R_\mu,
\label{mixing}
\end{eqnarray}
where
\begin{equation} 
\tan2\theta(q^2)=\frac{\sqrt{72m^2q^2}}{M^2_R-M^2_V}.
\label{theta}
\end{equation}
It means that $\rho$ and $\rho'$ mesons have both vector and
anomalous tensor couplings with quarks, while the axial-vector
mesons $a_1$ with quantum numbers $1^{++}$, which are assigned to
$A_\mu$, have only 
gauge-like minimal interactions and the axial-vector mesons $b_1$ 
with quantum numbers $1^{+-}$, which are assigned to
$B_\mu$, have only anomalous tensor interactions.

Diagonalization of (\ref{VR}) leads to 
relations for the physical masses of $\rho$,
$\rho'$ and $b_1$ mesons~\cite{PDG}:
$m_\rho=771.1\pm 0.9$ MeV,  $m_{\rho'}=1465\pm 25$ MeV and 
$m_{b_1}=1229.5\pm 3.2$~MeV:
\begin{eqnarray}
m^2_\rho+m^2_{\rho'}&=&M^2_V+m^2_{b_1}+6m^2\nonumber\\
m^2_\rho m^2_{\rho'}&=&M^2_V\left(m^2_{b_1}-12m^2\right),
\label{viete}
\end{eqnarray}
which define $M_V= 1034\pm 33$ MeV 
and the quark mass $m= 163\pm 7$ MeV. 
 
In general, the mixing angle $\theta$ depends on $q^2$ (\ref{theta}) 
and should be different at $\rho$- and $\rho'$-scale. However, the
denominator in (\ref{theta}) is smaller in comparison with the nominator
and $\tan2\theta(q^2)=|q|/(89^{+82}_{-75}~{\rm MeV})$ is big, 
which corresponds to almost maximal mixing $\theta\sim \pi/4$
with weak $q^2$-dependence.

If we suppose, that the effective four-fermion interactions of quarks
(\ref{4fermion}) could originate in QCD by gluon exchange in $1/N_c$
limit, it follows $M_V=M_A$~\cite{QCD}. Then from the first equation in 
(\ref{viete}) a remarkable relation among masses of the vector
mesons is obtained
\begin{equation}
m^2_\rho+m^2_{\rho'}=m^2_{a_1}+m^2_{b_1}.
\label{rel}
\end{equation}
This leads to a little bit smaller value for the mass of $a_1$ meson
$m_{a_1}= 1109\pm 37$ MeV ($2.2\sigma$ below PDG estimation).  

\section{Longitudinal and transverse polarizations of the vector mesons}

The longitudinal and transverse polarizations of the (axial-)vector mesons
are defined by corresponding couplings $f$ from the matrix elements
\begin{eqnarray}
\langle0|\bar{\psi}\gamma_\mu\gamma^5\psi|A(q,\lambda)\rangle&=&
m_A f_A e^\lambda_\mu\nonumber\\
\langle0|\bar{\psi}\sigma_{\mu\nu}\psi|A(q,\lambda)\rangle&=&
i f^T_A\varepsilon_{\mu\nu\alpha\beta}e^\lambda_\alpha q_\beta
\label{axial}\\
\langle0|\bar{\psi}\gamma_\mu\psi|V(q,\lambda)\rangle&=&
m_V f_V e^\lambda_\mu\nonumber\\
\langle0|\bar{\psi}\sigma_{\mu\nu}\psi|V(q,\lambda)\rangle&=&
i f^T_V\left(e^\lambda_\mu q_\nu - e^\lambda_\nu q_\mu\right),
\label{vector}
\end{eqnarray}
where $e^\lambda_\mu$ is the polarization vector of a spin-1 meson.
Then, using the meson-fermion couplings (\ref{boson}) and taking into account
the mixing after the symmetry breaking (\ref{mixing}), we can express
the couplings $f$ in terms of the parameters of our model:
\begin{eqnarray}
f_{a_1}&=&\frac{m_{a_1}}{g_A},\hspace{2.5cm} f^T_{a_1}=0;
\label{a1}\\
f_{b_1}&=&0,\hspace{3cm} f^T_{b_1}=\frac{m_{b_1}}{g_B};
\label{b1}\\
f_\rho&=&\frac{m_\rho\cos\theta(m_\rho)+\sqrt{18}m\sin\theta(m_\rho)}{g_V},
\nonumber\\
f^T_\rho&=&\frac{m_\rho\sin\theta(m_\rho)+\sqrt{18}m\cos\theta(m_\rho)}{g_R};
\label{rho}\\
f_{\rho'}&=&\frac{-m_{\rho'}\sin\theta(m_{\rho'})
+\sqrt{18}m\cos\theta(m_{\rho'})}{g_V},
\nonumber\\
f^T_{\rho'}&=&\frac{m_{\rho'}\cos\theta(m_{\rho'})
-\sqrt{18}m\sin\theta(m_{\rho'})}{g_R};
\label{rho'}
\end{eqnarray}
where $g_a$ obey the relations (\ref{ga}).

We can make immediately simple qualitative predictions, 
using the interesting fact that the solution of the system (\ref{viete})
is very close to the unique solution when the mixing angle $\theta$ equals
$\pi/4$ and does not depend on $q^2$. This happens when $M^2_R=M^2_V$.
In this case some valuable mass relations, besides (\ref{rel}), have place
\begin{eqnarray}
m_{\rho'}&=&m_\rho+\sqrt{18}m\nonumber\\
3m^2_{b_1}&=&2m^2_{\rho'}-m_\rho m_{\rho'}+2m^2_\rho\nonumber\\
3m^2_{a_1}&=&m^2_{\rho'}+m_\rho m_{\rho'}+m^2_\rho.
\label{unique}
\end{eqnarray}
The first two relations based only on the suggestion of maximal mixing, while
the last one requires of the additional constraint $M_V=M_A$.
Using more precise mass values for $\rho$ and $b_1$ mesons~\cite{PDG},
we can predict
other masses based solely on this ansatz: $m_{a_1}= 1155.1\pm 2.7$ MeV,
$m_{\rho'}= 1500.5\pm 4.8$~MeV and $m= 171.9\pm 1.3$ MeV.

It is interesting to note that the second relation can be applied
to the spin-1 isosinglets $I=0$ $\omega(782)$, $\omega'(1420)$, $h_1(1170)$
and $\phi(1020)$, $\phi'(1680)$, $h_1(1380)$
which are almost pure $(u\bar{u}+d\bar{d})/\sqrt{2}$ and $s\bar{s}$
states correspondingly. In the first case the mass relation among
$\omega$'s and $h_1$ is fair satisfied. In the second case we can confirm
existence of $h_1(1380)$ state which is omitted from summary table.
Our prediction for the mass $m_{h_1(1380)}=1415\pm 13$ MeV agrees
with PDG average $m_{h_1(1380)}=1386\pm 19$ MeV.

In the case of the maximal mixing
the eqs.\ (\ref{rho}) and (\ref{rho'}) can be rewritten 
in a compact form
\begin{eqnarray}
f_\rho&=&\frac{m_{\rho'}}{\sqrt{2}g_V},\hspace{1.5cm} 
f^T_\rho=\frac{m_{\rho'}}{\sqrt{2}g_R};
\label{rho0}\\
f_{\rho'}&=&-\frac{m_\rho}{\sqrt{2}g_V},\hspace{1.2cm} 
f^T_{\rho'}=\frac{m_\rho}{\sqrt{2}g_R};
\label{rho'0}
\end{eqnarray}
As far as $g_R=\sqrt{2}g_V$ eq.\ (\ref{rho0}) 
leads directly to a prediction for the ratio 
$f^T_\rho/f_\rho=1/\sqrt{2}\approx .707$, which is in perfect agreement
with the latest lattice calculations~\cite{lattice}. The analogous ratio
for the $\rho'$ meson should have the same value and an opposite sign
$f^T_{\rho'}/f_{\rho'}=-1/\sqrt{2}$. Unfortunately, there are no lattice
calculations for the $\rho'$ meson yet. 

The another reliable method, giving the information about
the matrix elements, is the QCD sum rules~\cite{SVZ}.
If one accepts the experimental value for the vector coupling
$f_\rho=208\pm 10$~MeV~\cite{lattice} 
the QCD sum rules give compatible with lattice calculation 
result $f^T_\rho=160\pm 10$~MeV~\cite{BB}. 
It is interesting to note that the consideration of the correlation
function of the tensor with the vector current proves
that the relative sign of $f^T_\rho$ and $f_\rho$ is positive~\cite{CZ}.
Moreover, it was shown~\cite{BK} that the contribution of $\rho'$
meson to this correlation function is negative and
$m_\rho f_\rho f^T_\rho \simeq -2 m_{\rho'} f_{\rho'} f^T_{\rho'}$.
That is in a good accordance with our predictions 
(\ref{rho0},\ref{rho'0}), 
since $m_{\rho'}\simeq 2 m_\rho$.

The derivation of the
transverse couplings within the QCD sum rules framework has been re-estimated
in~\cite{BM}: $f^T_\rho=157\pm 5$~MeV and $f^T_{b_1}=184\pm 5$~MeV,
confirming the results of~\cite{BB}. In addition the $\rho'$-transverse 
coupling $f^T_{\rho'}=140\pm 5$~MeV has been evaluated, which is,
however, in contradiction with the superconvergence relation
\begin{equation}
\left(f^T_\rho\right)^2+\left(f^T_{\rho'}\right)^2=
\left(f^T_{b_1}\right)^2
\label{super}
\end{equation}
from~\cite{CS}.

It should be compared with our predictions for the same couplings:
$f^T_\rho=(0.703^{+0.004}_{-0.007})f_\rho=146\pm 7$~MeV,
$f^T_{b_1}=(0.839^{+0.017}_{-0.015})f_\rho=175\pm 9$~MeV and
$f^T_{\rho'}=(0.405^{+0.040}_{-0.034})f_\rho=84\pm 9$~MeV. The first two
of them are in good agreement with all QCD sum rules results, while
the last one for the $\rho'$ meson is in a good accordance with the
relation (\ref{super}), but it is in noticeable disagreement with~\cite{BM}.

It is worth noticing that the anomalous dimension of the tensor current 
is not zero and the couplings $f^T(\mu)$ are scale dependent.
Our values are systematically lower than QCD
estimations made at the renormalization scale $\mu=1$~GeV
and closer to lattice calculations at \mbox{$\mu=2$~GeV}.

\section{Conclusion} 

In this letter we followed an approach based on the extended NJL quark model, 
describing all low-lying meson resonances. 
Totally there are 16 degrees of freedom
for the one-flavor quark-antiquark meson excitations:
scalar, pseudoscalar, vector, axial-vector and antisymmetric tensor.
They have corresponding Yukawa interactions with quarks.
All these excitations are assigned to physical meson states 
$\sigma$, $\pi$, $\rho$, $a_1$, $\rho'$ and $b_1$.
The (pseudo)scalar sector was already well studied and introduction 
of the tensor bosons does not bring anything new there.
Hence, we have concentrated on the spin-1 meson sector mainly. 
 
Simultaneous description of all these states in the framework of NJL
model leads to interesting predictions like new mass formulas  
and relations among meson coupling constants.
All these relations are in agreement with present experimental data
and the numerical calculations on the lattice and the QCD sum rules.

Other interesting property which follows immediately from our approach
is the dual nature of $\rho$ and $\rho'$ mesons. 
They have both vector and tensor couplings with quarks.
The new insight on this phenomenon is the suggestion that 
there exist two different vector particles with the same quantum numbers,
which interact differently with quarks. One of them has only
gauge-like vector interactions and the other one has only {\it anomalous}
tensor interactions. After the spontaneous chiral symmetry breaking
they are mixed, producing the physical $\rho$ and $\rho'$ meson states.
From the hadron phenomenology point of view
this suggestion does not seem unnatural, because the axial-vector  
mesons $a_1$ and $b_1$, due to their different quantum numbers,
exist as pure states, which have correspondingly only gauge-like couplings
and tensor couplings with quarks.

The above consideration means that in Nature 
there exist two different vector
particles with respect to their interactions with matter.
However, they are just composite quark-antiquark states. 
The question, if there are new fundamental vector particles,
will be probably answered at the future colliders and especially at LHC.

\vspace{-0.5cm}
\section*{Acknowledgments}

I would like to thank Professor V. Braun for useful comments. 
During preparation of this letter I learned about premature death 
of Professor Ian Kogan. All we miss him very much.
I acknowledge the warm hospitality of IPNL 
and especially Professors S. Katsanevas and Y.~D\'eclais.


\end{document}